\documentclass[prb,twocolumn]{revtex4-1}
\usepackage{amsmath}  
\usepackage{amsfonts} 
\usepackage{graphicx} 
\usepackage{graphicx}
\usepackage{dcolumn}
\usepackage{bm}
\usepackage{fancyhdr}
\usepackage{natbib}
\usepackage{color}

\usepackage{makecell}

\begin{document}
\title{Diffraction effects in mechanically chopped laser pulses}
\author{Samridhi Gambhir and Mandip Singh}
\email{mandip@iisermohali.ac.in}
\affiliation{Department of Physical Sciences,
\\Indian Institute of Science Education and Research Mohali, 140306,
India.}
\begin{abstract}
A mechanical beam chopper consists of a rotating disc of regularly spaced wide slits which allow light to pass through them. A continuous light beam, after passing through the rotating disc, is switched-on and switched-off periodically, and a series of optical pulses are produced. The intensity of each pulse is expected to rise and fall smoothly with time. However, a careful study has revealed that the edges of mechanically chopped laser light pulses consist of periodic intensity undulations which can be detected with a photo detector. It has been shown in this paper that the intensity undulations in mechanically chopped laser pulses are produced by diffraction of light from the rotating disc and a detailed explanation of the intensity undulations is given. The experiment provides an efficient method to capture a one dimensional diffraction profile of light from a straight sharp-edge in the time domain. In addition, the experiment accurately measured wavelengths of three different laser beams from the undulations in mechanically chopped laser light pulses.   

\end{abstract}
\maketitle

\section{Introduction}
Diffraction of light originates from the wave nature of light \cite{jenkins, born, hecht}. Many experiments and techniques are based on the diffraction \cite{im1, im2, im3, scholten} however, diffraction can also produce artifacts. For example, in imaging experiments with coherent light the unavoidable diffraction can produce patterns in the image field \cite{atom}. Such a diffraction patten can be produced by an obstacle located in the path of a light beam. A  dust particle or a debris sticking on optical components can also result a fringe pattern which can distort the image field. A similar situation of an unavoidable diffraction effect appears in a mechanically chopped laser beam which is unnoticed in many experiments. An ideal beam chopper switches-on and switches-off the light suddenly. However, diffraction of light from the blades of a beam chopper disc can produce undulatory spikes at the leading and the trailing edges of light pulses. Intensity variations of the chopped light pulses can be detected with a photo detector. The photo detector output signal shows a pattern of decaying periodic undulations at the edges of the pulses which resembles to a ringing artifact of electrical pulses in electronic circuits \cite{johnson, gibb}. However, the nature of undulatory pattern in mechanically chopped light pulses is different from the ringing artifact, where the latter is caused by distortion of the frequency spectrum of electrical pulses and the former is caused by the spatial diffraction of light beam chopped in time domain. The undulatory pattern at edges of a chopped pulse is observed if the photo detector size is kept small  or if a stationary pin hole (or a single slit) is placed in front of a large area stationary photo detector to reduce the exposed area of photo detector. Diffraction of light from a stationary straight sharp-edge has been studied and demonstrated in many different experiments \cite{C,B,A,F,a1}. In all such experiments, a straight edge is kept stationary. In this paper, a detailed study of diffraction of a laser light beam from moving straight sharp-edges of a mechanical light beam chopper disc is presented. This experiment provides a new method, to acquire a one dimensional diffraction profile on an oscilloscope in time domain very quickly and to accurately measure an unknown wavelength of a laser beam.
 \begin{center}
\begin{figure}
\begin{center}
\includegraphics[scale= 0.215]{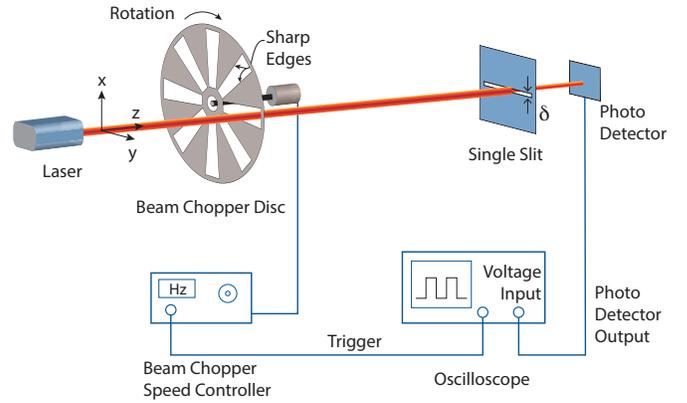}
\caption{\label{fig1} \emph{A schematic diagram of a laser light beam chopper experiment. Light is diffracted from moving sharp edges of a beam chopper disc. As the disc is rotated the diffraction pattern is shifted with time. A single slit is placed in-front of a photo detector to select a small part of the moving diffraction pattern at an instant of time. Photo detector output voltage signal in time domain corresponds to a one dimensional diffraction profile of the light beam from a moving straight sharp-edge.}}
\end{center}
\end{figure}
\end{center}

\section{Experiment}
 \begin{center}
\begin{figure*}
\begin{center}
\includegraphics[scale= 0.36]{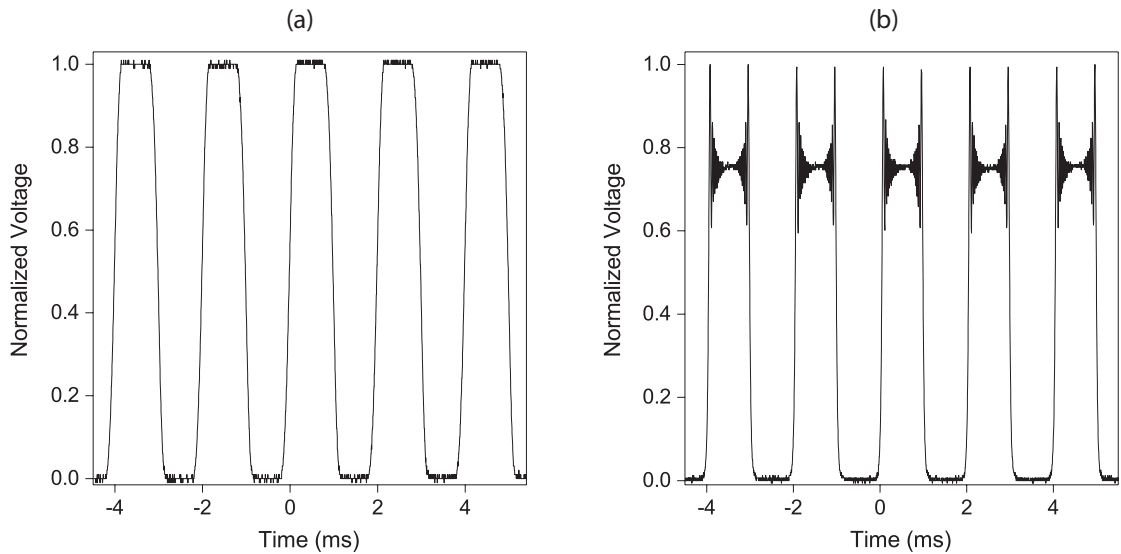}
\caption{\label{fig2} \emph{(a) Photo detector output voltage with time. In this plot the chopped laser light pulses are incident on a large area photo detector. (b) If a single slit of narrow slit width is placed in-front of a large area photo detector then spikes appear at the leading and the trailing edges of a photo detector output voltage signal.}}
\end{center}
\end{figure*}
\end{center}
A schematic of a laser light beam chopping experimental set-up is shown Fig.~\ref{fig1}, where a laser beam is passed through a rotating disc of a mechanical beam chopper. The chopped light pulses are detected by a large area fast response photo detector whose output voltage is linearly proportional to intensity of incident light. A laser beam propagating along $z$-axis is incident perpendicular to the plane of the disc and the plane of the disc is kept parallel to $x$-$y$ plane. The disc is aligned such that the straight edges of the disc slot become parallel to $y$-axis at the point of intersection of centre of light beam and the disc. Linear velocity of disc at a point of intersection is in a negative $x$-direction. The radial distance of the point of intersection can be adjusted precisely. Photo detector output is connected to a digital oscilloscope which is synchronized by a trigger signal generated by the beam chopper speed controller. Slots of the chopper disc, through which the light passes, are equally spaced and the disc is rotating with a constant angular velocity.  If the chopped laser beam pulses are detected by a large area photo detector in absence of a single slit (or a pin hole) in front of the photo detector then a smooth rise and fall of intensity of light is observed as shown in Fig.~\ref{fig2} (a), which is a plot of the photo detector output voltage with time. This experiment is performed with a Helium-Neon (He-Ne) laser light of wavelength 633~nm. A finite width of laser beam along transverse plane leads to a smooth rise and fall in the intensity of chopped light pulses. 

If light pulses are detected by a small area photo detector or if a single slit is placed in front of a large area photo detector then sudden jumps of intensity at the leading and the trailing edges of the chopped light pulses are observed as shown in Fig.~\ref{fig2} (b.) This is the plot of  photo detector output voltage with time. A single slit is located in a plane parallel to $x$-$y$ plane and its width along $x$-axis is $\delta=$~0.1~mm, which is much smaller than fringe separation, and slit length along $y$-axis is 15~mm. Slit width $\delta$ is the effective photo detector size therefore, by varying the slit width the effective detector size is varied. Laser beam is expanded and collimated by placing two convex lenses in the beam path. A rectangular aperture, of width about 1~mm in the radial direction and length about 20~mm in the perpendicular direction, is placed in the path of expanded beam.  This configuration produces a beam extension about 1~mm in the radial direction and a beam extension about 5~mm (full width at a half maximum of intensity) in a direction perpendicular to the radial direction just before the beam chopper disc. Laser beam is expanded to increase the number of observed fringes. A distance between the disc and a single slit is 1.74~m (slit is placed very close to photo detector) and the radial distance of a point of intersection (along $y$-axis) \emph{i.e.} where a laser beam center intersects with the disc is 47~mm. At this point a linear velocity of disc is -14.78 m/sec along a negative $x$-direction.
\begin{center}
\begin{figure*}
\begin{center}
\includegraphics[scale= 0.36]{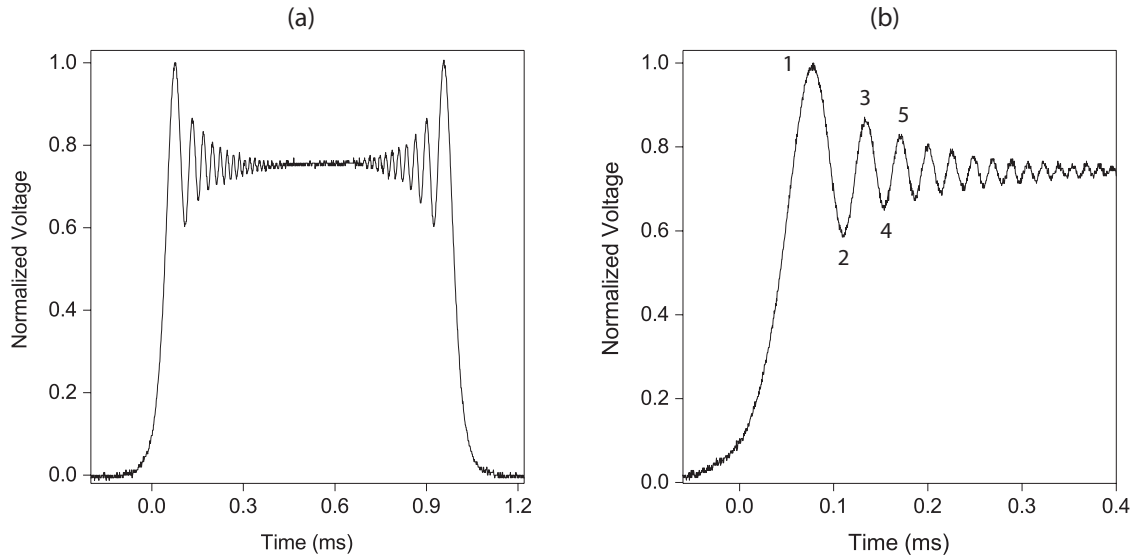}
\caption{\label{fig3} \emph{(a) Spikes consist of undulations at the leading and the trailing edges of each pulse. (b) A magnified capture of the photo detector voltage signal around the leading edge corresponds to a diffraction profile of light from a single straight sharp-edge of the beam chopper disc. An inverted profile at the trailing edge shown in (a) corresponds  to a diffraction profile of light from the second edge of the same slot of the rotating beam chopper disc. This diffraction profile is obtained with a He-Ne laser light of wavelength 633~nm.}}
\end{center}
\end{figure*}
\end{center}

 If a semi-infinite straight sharp-edge is placed in the path of a plane wave, the wave gets diffracted which results in a nonzero wave amplitude in the geometrical shadow of the edge and formation of a fringe pattern. Fringe visibility is predominant near the boundary of a geometrical shadow. Intensity of fringes and a separation between consecutive fringes decrease with distance from the boundary. In the context of experiment presented in this paper the undulatory spikes in mechanically chopped laser pulses originates due to diffraction of light from sharp-edges of the chopper disc slots through which the light beam is passed. As a moving edge of a slot of the chopper disc comes in the path of the laser beam the light gets diffracted from the straight edge of the slot. Since the straight edges are moving therefore, the diffraction pattern is displaced in the direction of linear velocity of the chopper disc at a point where the laser beam intersects the chopper disc. A moving beam chopper disc produces a moving diffraction pattern. Since a large area photo detector collects almost a complete range of observable diffraction pattern at a time therefore, the undulatory pattern is not observed at the output of the photo detector. However, if a single slit is placed in front of a large area photo detector as shown in Fig.~\ref{fig1} then the light from a small portion of the moving diffraction pattern is detected at a time and the diffraction pattern is resolved in time domain. As a sharp-edge of a disc slot is moved the diffraction pattern is displaced on the stationary single slit and the intensity transmitted through the slit is measured in real time point by point. Photo detector output corresponds to a high resolution one dimensional diffraction profile in real time.
 
In a conventional diffraction experiment, a photo detector is displaced to measure the intensity at different points in the diffraction pattern. In this paper, photo detector is kept stationary while the diffraction pattern is displaced.  This method leads to a real time and quick high resolution measurements of the diffraction pattern. In this paper a single narrow slit is preferred over a pin hole to increase signal to noise ratio.  The output voltage of the photo detector for a single pulse is shown in Fig.~\ref{fig3} (a) which shows that the spikes consist of undulations at the leading and the trailing edges of a light pulse. A magnified view of the undulatory pattern at the leading edge of the pulse is shown in Fig.~\ref{fig3} (b). This undulatory pattern corresponds to the diffraction pattern of light from a single straight sharp-edge. If a slit located in front of the photo detector is rotated in a plane parallel to the $x$-$y$ plane such that its length is aligned parallel to the $x$-axis then the undulatory pattern disappears. In this case the fringes are moving parallel to the length of the slit and total intensity of fringe pattern is detected at an instant of time. Therefore, output voltage waveform of a photo detector is similar to voltage signal shown in Fig.~\ref{fig2} (a).

\subsection{Theory and wavelength measurements}
\begin{center}
\begin{figure}
\begin{center}
\includegraphics[scale= 0.32]{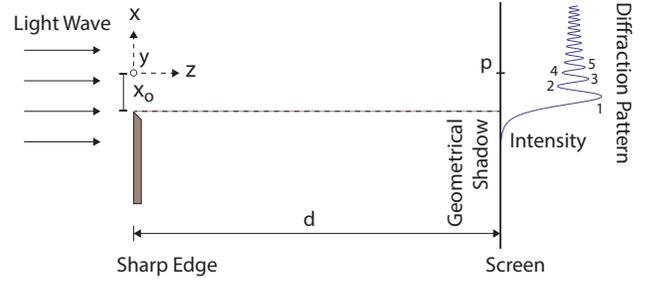}
\caption{\label{fig4} \emph{ A schematic diagram showing a diffraction of light from a straight sharp-edge.}}
\end{center}
\end{figure}
\end{center}

Consider a semi-infinite straight sharp-edge located in $x$-$y$ plane as shown in Fig.~\ref{fig4}, where a light wave of planar wavefront is propagating along $z$-axis. A third axis ($y$) is perpendicular to plane of paper. A semi-infinite planar sharp-edge located in $x$-$y$ plane blocks the propagation of light for $x \leq x_{o}$ in $x$-$y$ plane. According to Huygens-Fresnel principle, each point on a wavefront is considered as a point source emanating a spherical wave. Spherical waves from all the points from a previous (primary) wavefront are superimposed and a new (secondary) wavefront is formed at a new displaced location. This process continues and governs the evolution of wave amplitude on the progressive wavefronts. A planar wavefront evolves to a planar wavefront however, if any part of the primary wavefront is blocked then the contribution of spherical waves emanating from the blocked region to the superposition becomes zero. As a result a complete cancellation of wave amplitudes does not happen in the region of geometrical shadow which leads to a diffraction of a wave  \emph{i.e.} a non zero wave amplitude in the geometrical shadow. In present situation of a semi-infinite straight sharp-edge the intensity of light wave exhibits an undulatory pattern, which is predominant close to the geometrical shadow with a non zero intensity in the geometrical shadow, as shown in Fig.~\ref{fig4}. 

To analyze this, consider a stationary point $p$ in the $x$-$z$ plane on the screen with coordinates (0,0,$d$), where $d$ is the distance between a screen and a straight sharp-edge. Total electric field at a point $p$ is a superposition of spherical waves amplitudes emanated from the unobstructed region of $x$-$y$ plane. Consider an infinitesimal area element $\mathrm{d}x\mathrm{d}y$ located at a point of coordinates ($x$,$y$, 0) in $x$-$y$ plane. Electric field at $p$ of spherical waves emitted from the area element is $\mathrm{d}E = c_{o} (e^{i(kr-\omega t)}/r ) \mathrm{d}x\mathrm{d}y$, where $\omega$ is angular frequency of light, $k=2\pi/ \lambda$ is a propagation constant corresponding to wavelength $\lambda$, $c_{o}$ is a constant of proportionality and $r$ is the distance of a point $p$ on the screen from the area element. For a normal incidence of a planar wavefront on a semi-infinite straight sharp-edge the total electric field at $p$ can be written as   

\begin{equation}
\label{eqn1}
  E(x_{o},t)=c_{o} \int_{x_{o}}^{\infty} \int_{-\infty}^{\infty} \frac{e^{i(kr-\omega t)}}{r} {d}x{d}y 
\end{equation}
At a point $p$, the contribution  of spherical waves to the integral in Eq.~\ref{eqn1} diminishes as $r$ increases and integral saturates as a limit of $r$ increases \emph{i.e.} a region $(x^{2}+y^{2})^{1/2} \gg d$, where $r$ can be written as $r=(d^{2}+x^{2}+y^{2})^{1/2}$. Therefore, an obliquity factor is considered to be one. A first order Binomial expansion of $r$ for $d\gg(x^{2}+y^{2})^{1/2}$ is $r\approx d +(x^{2}+y^{2})/2 d$. Therefore, Eq.~\ref{eqn1} can be written as
\begin{equation}
\label{eqn2}
  E(u_{o},t)\approx \frac{c_{o} \lambda e^{i(kd-\omega t)}}{2} \int_{u_{o}}^{\infty} 
 e^{i\pi u^{2}/2}{d}u 
  \int_{-\infty}^{\infty} e^{i\pi v^{2}/2} {d}v 
\end{equation}
Where, $u=(2/\lambda d)^{1/2}x$ and $v=(2/\lambda d)^{1/2}y$ and $u_{o}=(2/\lambda d)^{1/2}x_{o}$. 
Integrals in Eq.~\ref{eqn2} can be expressed in the form of standard Fresnel integrals, $C(w) =  \int_{0}^{w} \cos(\pi w'^{2}/2) {d}w'$ and $S(w) =  \int_{0}^{w} \sin(\pi w'^{2}/2) {d}w'$ by applying $ \int_{0}^{w} e^{i\pi w'^{2}/2} {d}w'= C(w)+iS(w)$. Therefore, intensity $I=E E^{*}$ of light wave at point $p$ can be evaluated as
\begin{equation}
\label{eqn3}
 I (u_{o}) \approx \frac{I_{o}}{2}([\frac{1}{2}-C(u_{o})]^{2}+[\frac{1}{2}-S(u_{o})]^{2})
\end{equation}
This expression represents a diffraction profile of a semi-infinite planar sharp-edge, where $I_{o}$ is intensity at a point $p$ in absence of a semi-infinite planar sharp-edge. If a sharp edged semi-infinite plane is moved with velocity $v$ along $x$-axis then $x_{o}=vt$ at an instant of time $t$. In this way the diffraction pattern is displaced and intensity variation with time at a fixed point $p$ corresponds to a diffraction profile from a semi-infinite straight sharp-edge in time domain as shown in Fig.~\ref{fig3} (b). A diffraction pattern obtained by moving the sharp-edge for a stationary location of $p$ is equivalent to a diffraction pattern obtained by moving $p$ in the opposite direction while keeping the sharp-edge stationary.

 \begin{center}
\begin{figure*}
\begin{center}
\includegraphics[scale= 0.37]{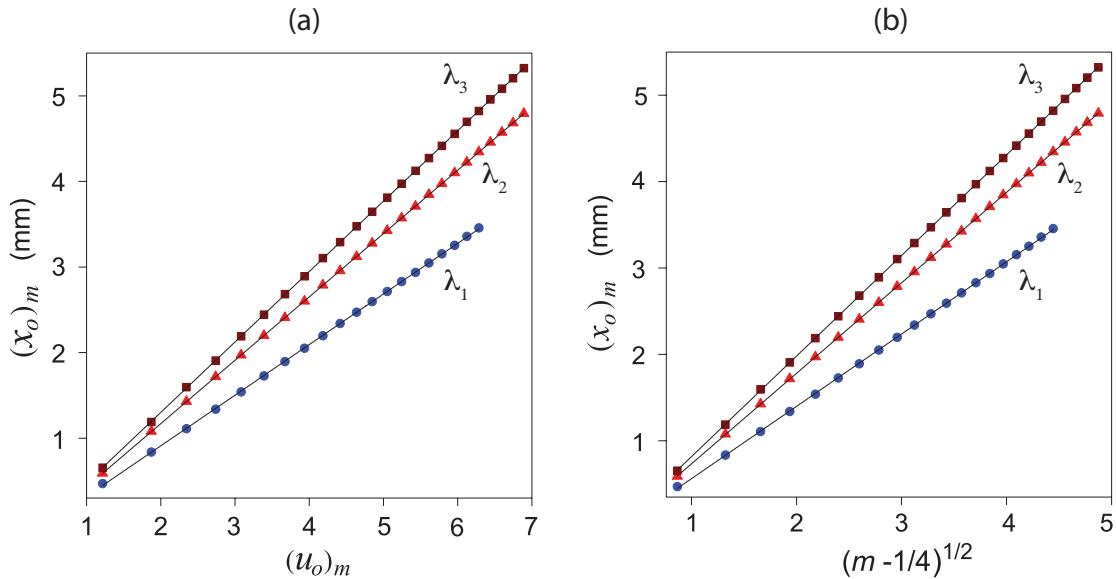}
\caption{\label{fig5} \emph{(a)  A plot of extremum point locations $(x_{o})_{m}$ corresponding to calculated values of $(u_{o})_{m}$ for three different wavelengths (method-I). (b) A plot of extremum point locations $(x_{o})_{m}$ corresponding to $(m-1/4)^{1/2}$ for three different wavelengths (method-II).}}
\end{center}
\end{figure*}
\end{center}

An unknown wavelength of laser light is measured from the diffraction profile of a beam chopper sharp-edge as shown in Fig.~\ref{fig3} (b). In  Fig.~\ref{fig3} (b) and Fig.~\ref{fig4} the locations of extremum points in the diffraction profile are labeled by nonzero positive integers $m$, where $m=1$ corresponds to a first extremum point (first maxima close to the geometrical shadow), $m=2$ corresponds to a next extremum point (first minima close to the geometrical shadow) and so on. Extremum points $(u_{o})_{m}$ of diffraction function $I (u_{o})$ given in Eq.~\ref{eqn3} are evaluated and the locations of extremum points $t_m$ of a measured diffraction pattern shown in Fig.~\ref{fig3} (b) are evaluated in temporal domain and transformed to a spatial domain  by using $(x_{o})_{m}= v t_{m}$.  Wavelength is evaluated from $\lambda= 2 (x_{o})_{m}^{2}/ (u_{o})_{m}^{2}d$ \emph{i.e.} from relation between variables $u$ and $x$. Laser beam chopper diffraction experiment is performed with three different lasers beams of different wavelengths and a diffraction pattern corresponding to each wavelength is obtained. A plot between $(u_{o})_{m}$ and $(x_{o})_{m}$ is shown in Fig.~\ref{fig5} (a) for three different wave lengths $\lambda_{1}$,  $\lambda_{2}$  and $\lambda_{3}$, where each point on a line in the plot represents an extremum point in the diffraction profile corresponding to a wavelength. A positive sign for values of $(u_{o})_{m}$ and $(x_{o})_{m}$ is considered in plots shown in Fig.~\ref{fig5}. Wavelength of laser light is evaluated from the slope ($s$) of a best fit function \emph{i.e.} $\lambda = 2 s^{2}/d $. For comparison and error estimation the wavelength of three different laser beams are separately measured with an optical spectrum analyzer.  Wavelength measurements are shown in Table.~\ref{table1} along with error estimation by comparing these wavelength measurements with wavelength measurements performed with an optical spectrum analyzer. This is a method-I.

\begin{table}
\caption{Wavelength measurements with method-I} \label{table1}
\begin{center}
\begin{tabular}{  |c || c | c | r|  }
  \hline
  \makecell{Wavelength \\ (nm)} & \makecell{Measured from \\ diffraction\\ profile} & \makecell {Measured with an \\ optical spectrum\\ analyzer} & \makecell{Error\\(\%)}  \\
  \hline 
  $\lambda_{1}$  & 403.1  & 405.4  & 0.57  \\
  $\lambda_{2}$  & 631.8  & 633.1  & 0.20  \\
  $\lambda_{3}$  & 780.5  & 780.9  & 0.05  \\
  \hline
\end{tabular}
\end{center}
\end{table}
\begin{table}
\caption{Wavelength measurements with method-II} \label{table2}
\begin{center}
\begin{tabular}{  |c || c | c | r|  }
  \hline
  \makecell{Wavelength \\ (nm)} & \makecell{Measured from \\ diffraction\\ profile} & \makecell {Measured with an \\ optical spectrum\\ analyzer} & \makecell{Error\\(\%)}  \\
  \hline 
  $\lambda_{1}$  & 403.5  & 405.4  & 0.47  \\
  $\lambda_{2}$  & 632.3  & 633.1  & 0.13  \\
  $\lambda_{3}$  & 781.0  & 780.9  & 0.01  \\
  \hline
\end{tabular}
\end{center}
\end{table}

In a method-II, wavelengths of all three laser beams are evaluated from location of extremum points, which can be expressed as $(x_{o})_{m}=(d\lambda)^{1/2}(m-\frac{1}{4})^{1/2}$ \cite{D,E}. This expression is derived by superimposing a plane wave propagating in $z$-direction with a spherical wave emitted by an imaginary source located on an edge of a semi-infinite planar opaque obstacle.  Wavelengths of three different laser beams are then calculated from slope of the best fit plot of $(x_{o})_{m}$  and $(m-\frac{1}{4})^{1/2}$ as shown in Fig.~\ref{fig5} (b). Measurements of wavelengths, of three different laser beams, with method-II are shown in Table.~\ref{table2}. Error of measurement is estimated by comparing a wavelength with a corresponding wavelength measurement performed with an optical spectrum analyzer.

\begin{center}
\begin{figure*}
\begin{center}
\includegraphics[scale= 0.32]{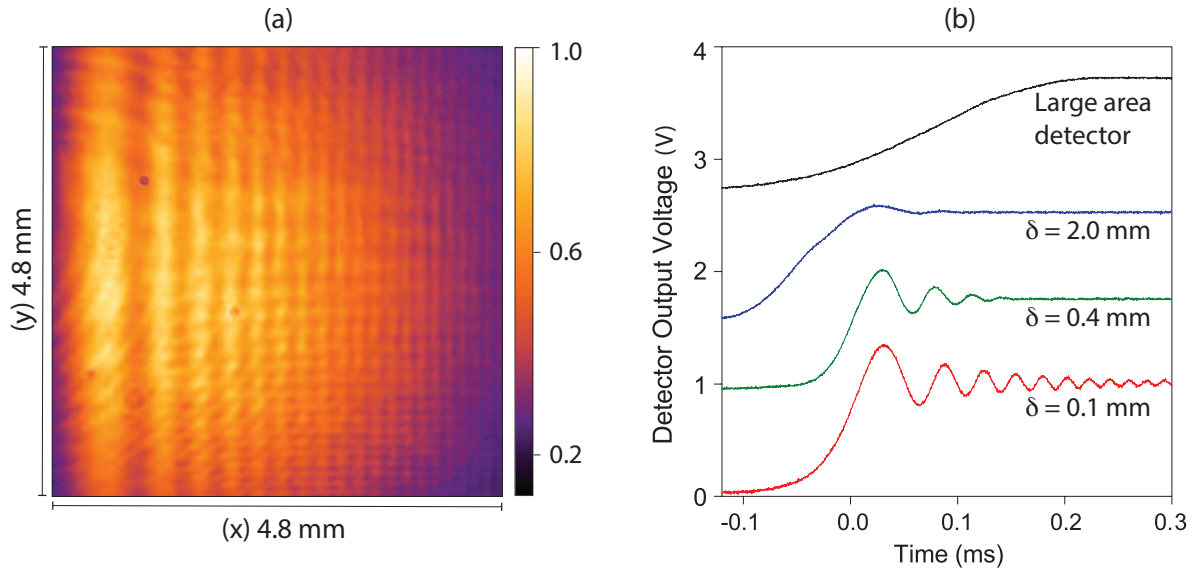}
\caption{\label{fig6} \emph{ (a) A direct capture of a diffraction pattern (normalized intensity) of light from a sharp-edge of a stationary disc of an optical beam chopper.  (b) Effect of slit width $\delta$ on the undulatory spikes observed in a photo detector output signal.}}
\end{center}
\end{figure*}
\end{center}

In above measurements, a diffraction profile is captured in time domain. In another experiment, a diffraction pattern of a He-Ne laser light beam from a stationary sharp-edge of a laser beam chopper is directly captured with a complimentary metal-oxide semiconductor (CMOS) camera as shown in Fig.~\ref{fig6} (a). Laser beam is expanded up to $6$~mm to capture more than twenty observable fringes and a sharp-edge is aligned parallel to $y$-axis. In this experiment, a distance between a sharp-edge of a laser beam chopper and a camera sensor is $1$~m. Wavelength of He-Ne laser is evaluated from the diffraction pattern captured with camera by applying method-I and wavelength is compared with a wavelength measurement performed with an optical spectrum analyzer. The wavelength measured from the diffraction profile is $632.4$~nm and a  wavelength measured with an optical spectrum analyzer is $633.1$~nm. A percentage error in the measurement of wavelength calculated from a direct capture of diffraction pattern is $0.1~\%$.  

\subsection{Effect of size of photo detector sensor}

Diffraction of a laser beam from sharp edges of a laser beam chopper is observed if the detector size is much smaller than the fringe separation at the detector location. In experiments shown in this paper, a single slit of width $\delta$ is placed in front of a large area detector as shown in  Fig.~\ref{fig1}. The slit width $\delta$ is regarded as an effective detector size. If slit width is comparable or larger than the fringe separation then diffraction pattern in mechanically chopped laser beam pulses cannot be resolved since light from various portions of a fringe will be collected by the detector at an instant of time. The effect of the width $\delta$ of the single slit on the undulatory pattern on the leading edge of the chopped light pulse observed with a photo detector is shown in Fig.~\ref{fig6} (b) for $\delta$ equal to $0.1$~mm, $0.4$~mm, $2.0$~mm and for a large area photo detector without the single slit, where diameter of photo detector sensor is about $10$~mm. The pattern gradually disappears as the effective detector size is increased.       

\section{Conclusions}
This paper has experimentally found a reason of intensity undulations in mechanically chopped laser light pulses. Diffraction of a laser beam from sharp edges of a beam chopper disc produces intensity undulations in time domain at the leading and the trailing edges of chopped light pulses. The chopped pulses are passed through a single slit and then these pulses are detected with a fast response photo detector. Photo detector output gives a real time measurement of a one dimensional diffraction profile of light beam from an edge of each slot of the beam chopper disc. Diffraction effects remain unnoticed in a photo detector output voltage signal if area of the photo detector is much larger than the fringe separation. Wavelengths of three different lasers are measured accurately from the undulatory diffraction pattern of chopped laser beam pulses. Measurement results are compared with measurements based on a direct capture of the diffraction pattern with a camera. The effect of photo detector size on the undulatory pattern is also shown. In this way, this paper has shown an experimental method to precisely capture a diffraction profile in the real time with a stationary photo detector and to accurately measure a wavelength of a laser beam.




\end{document}